\documentstyle[amssymb,aps,epsf,twocolumn]{revtex}

\def\be{\begin{equation}}
\def\ee{\end{equation}}
\def\bea{\begin{eqnarray}}
\def\eea{\end{eqnarray}}
\def\nn{\nonumber\\}

\def\r#1{(\ref{#1})}

\begin{document}
\title{A numerical method for detecting incommensurate correlations 
in the Heisenberg zigzag ladder}
\author{A. A. Aligia$^{a}$, C. D. Batista$^{a}$  and F.H.L. E\ss ler$^{b}$}
\address{$^{a}$Comisi\'{o}n Nacional de Energ{\'{\i }}a At\'{o}mica,\\
Centro At\'{o}mico Bariloche and Instituto Balseiro, 8400 S.C. de Bariloche,%
\\
Argentina\\
$^{b}$Department of Physics, Warwick University\\
Coventry, CV4 7AL\\
United Kingdom.}

\date{Received \today }
\maketitle

\begin{abstract}
We study two Heisenberg spin-1/2 chains coupled by a frustrating
``zigzag'' interaction. We are particularly interested in the regime 
of weak interchain coupling, which is difficult to analyse by either
numerical or analytical methods. Previous density matrix
renormalisation group (DMRG) studies of the isotropic model with open
boundary conditions and sizeable interchain coupling have established
the presence of incommensurate correlations and of a spectral
gap. By using twisted boundary conditions with arbitrary twist angle,
we are able to determine the incommensurabilities both in the isotropic 
case and in the presence of an exchange anisotropy by means of exact 
diagonalisation of relatively short finite chains of up to $24$ sites.
Using twisted boundary conditions results in a very smooth dependence
of the incommensurabilities on system size, which makes the
extrapolation to infinite systems significantly easier than for open or
periodic chains. 
\end{abstract}

\pacs{PACS Numbers: 75.10.Jm, 75.40.Gb}
\narrowtext
\section{Introduction}
In recent years several frustrated quasi one dimensional magnetic
compounds have been identified and studied experimentally
\cite{srcuo2,matsuda,radu,john}. In the one-dimensional phase, i.e. for
temperatures above the magnetic ordering transition, frustration 
is expected to lead to incommensurate correlations. Precisely how
frustration gives rise to incommensurabilitites for the extreme
``quantum'' case of spin $S=1/2$ is at present not well understood. 

A paradigm of a frustrated quantum magnet is the spin-1/2
Heisenberg antiferromagnetic chain with nearest neighbour exchange
$J_1$ and next-nearest neighbour exchange $J_2$. This model is
equivalent to a two-leg ladder (see Fig. 1), where the coupling along
(between) the legs of the ladder is equal to $J_2$ ($J_1$).

The Hamiltonian is given by:

\bea
H&=&\sum_{i}[J_{1}(S_{i}^{x}S_{i+1}^{x}+S_{i}^{y}S_{i+1}^{y}\Delta
S_{i}^{z}S_{i+1}^{z})\nn
&&+\sum_iJ_{2}(S_{i}^{x}S_{i+2}^{x}+S_{i}^{y}S_{i+2}^{y}+\Delta
S_{i}^{z}S_{i+2}^{z})],  \label{h}
\eea
where we have allowed for an exchange anisotropy $\Delta$.

\begin{figure}
\narrowtext
\epsfxsize=3.0truein
\vbox{\hskip 0.05truein \epsffile{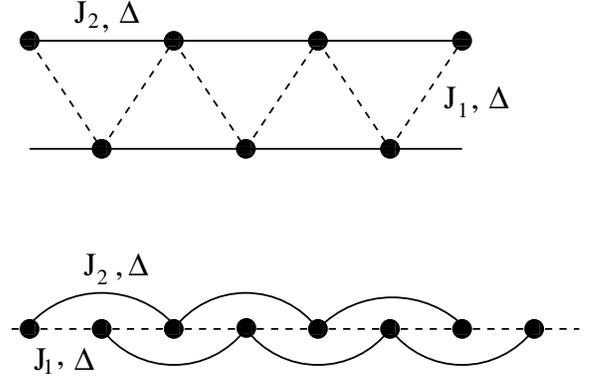}}
\medskip
\caption{Schematic representation of a zig-zag ladder (top) and
its equivalent chain with first and second nearest-neighbor exchange 
interactions (bottom).}
\end{figure}

The zigzag ladder model is believed to describe the quantum magnet
${\rm SrCuO_2}$ \cite{srcuo2,matsuda} above the magnetic ordering
transition. The ratio of exchange constants is estimated as
$|J_1/J_2|\approx 0.1-0.2$ \cite{matsuda}, so that the interchain
coupling is significantly smaller than the exchange along the legs of
the ladder.
A second well-studied material with zigzag structure is ${\rm
Cs_2CuCl_4}$ \cite{radu}. However, in ${\rm Cs_2CuCl_4}$ the chains
appear to be coupled in an entire plane and no pronounced ladder
structure is present.

The model \r{h} in the regime $J_2{>\atop\sim}J_1$ has been studied
previously using both numerical \cite{bur,whi} and field-theoretical
\cite{whi,NGE,allen,aen,chp} methods. The main DMRG results for the
spin rotationally symmetric case ($\Delta=1$) and not too large values
of $J_2/J_1$ are the following \cite{whi}
\begin{itemize}
\item{}
The ground state is doubly degenerate and is characterized by a nonzero
dimerisation $d=\langle
\vec{S}_{2n}\cdot\left(\vec{S}_{2n-1}-\vec{S}_{2n+1}\right)\rangle$.
\item{}
The equal-time correlation function $\langle \vec{S}_n\cdot
\vec{S}_1\rangle$ exhibits an oscillating exponential decay at large
spatial separations. The characteristic angle associated with these
oscillations, i.e. the incommensurability, is related to the
correlation length $\xi$ by
\be
\frac{\theta}{\pi/2}-1=\frac{1}{2\xi}.
\ee
This connection of incommensurability and correlation length also fits
into the picture emerging from the renormalisation group analysis
(valid in the limit $J_2\gg J_1$) of \cite{NGE}, which yields the {\sl
simultaneous} divergence, with fixed ratio, of the related coupling
constants.
\end{itemize}

The regime $J_2\gg J_1$ is very difficult to analyse numerically as
both the dimerisation and the incommensurability become very small.
In particular, the DMRG analysis of \cite{whi} did not consider this
regime.
At the same time the field theory studies of \cite{NGE} suggest that
for $J_2\gg J_1$ and $|\Delta|<1$ a different type of physics may emerge:
there are still incommensurate correlations and dimerisation, but in
addition there is also ``chiral'' order
\bea
&&\langle{S}^+_{2n}{S}^-_{2n+2}-{S}^-_{2n}{S}^+_{2n+2}\rangle\neq 0\
,\nn
&&\langle{S}^+_{2n}{S}^-_{2n-1}-{S}^-_{2n}{S}^+_{2n-1}\rangle\neq 0\ .
\eea
Such type of order is forbidden in the isotropic case $\Delta=1$.
It clearly would be interesting to numerically analyse whether or not
such a new phase indeed exists. Given the difficulties, mainly due to
finite-size effects, in accessing the relevant parameter regime by
numerical methods, we propose in essence to {\sl utilise} finite-size
effects to extract information on the incommensurability present in
the system. This is done by studying the effects of twisted boundary
conditions on the energy levels. Our main purpose is to establish the
viability of our method by carrying out exact diagonalisation of
finite clusters of up to 24 sites. In order to obtain definitive
results on the zigzag chain in the most interesting parameter regime,
larger systems need to be studied, possibly by implementing TBA into a
DMRG algorithm.

\section{The method}

We found the ground state of the system by Lanczos diagonalization of rings
of size $L=12$, 16, 20 and 24 sites in the subspaces of total spin
projection $S^{z}=0$, 1, for each total wave number $K$. We used twisted
boundary conditions (TBC) $S_{i+L}^{-}=e^{i\Phi }S_{i}^{-}$. This means that
each time a spin down traverses one particular link, it acquires a phase $%
e^{i\Phi }$ ($e^{-i\Phi })$ if it moves to the right (left). In the
fermionic representation, after a Wigner-Jordan transformation $%
S_{i}^{-}=c_{i}^{\dagger }\exp (i\pi \sum_{j<i}c_{j}^{\dagger }c_{j})$, and
a gauge transformation $c_{i}^{\dagger }=e^{i\Phi /L}f_{i}^{\dagger }$, the
problem becomes equivalent to a system of $N_{\downarrow }$ fermions ($f$)
on a ring threaded by a flux $\Phi $, where $%
N_{\downarrow }$ is the number of spins down. The advantage of the TBC for
our purposes is that the allowed total wave vectors are:

\begin{equation}
K_{n}(\Phi )=\frac{2\pi }{L}n+\frac{\Phi }{L}N_{\downarrow },  \label{ks}
\end{equation}
with $n$ integer. Thus, varying $\Phi $, we have access to a continuum of
possible wave vectors, even if we are working on a finite system \cite{tbc}.
As an example, the ground state of Eq. (\ref{h}) for $N_{\downarrow }=1$ is
known exactly. In the thermodynamic limit, for $1/4<\alpha =J_{2}/J_{1}<1/2$%
, the ground state is two fold degenerate with incommensurate wave vectors $%
K=\pm \arccos [-1/(4\alpha )]$ \cite{ger}. For small systems, these wave
vectors are only accessible for discrete particular values of $\alpha $ if
PBC are used. Instead, for any $\alpha $, the exact $K$'s and ground state
energies are reproduced, if the energy of a small ring is minimized  as a
function of flux  $\Phi $ and discrete wave number ($n$).

For $|\Delta |\eqslantless 1$, the transverse spin correlations
dominate at large distances. We denote the ground state of the system
by $|g\rangle$. It lies in the $S^{z}=0$ sector and its wave
vector for PBC (or TBC if the energy is minimized over $\Phi $) is
always $K_{0}=0$ or $\pi $. For the transverse spin correlations we
can write: 

\begin{equation}
C(l)=\langle g|S_{i}^{+}S_{i+l}^{-}|g\rangle =\frac{1}{L}\sum_{e,q}e^{-iql}|\langle
e|S_{q}^{-}|g\rangle |^{2},  \label{cf}
\end{equation}
where $S_{q}^{-}$ is the Fourier transform of $S_{l}^{-}$, and the sum
over $|e\rangle $ runs over all excited states.

By symmetry, for each $q$, only excited states with $S^z=-1$ and $%
K_{1}=K_{0}+q$ have nonvanishing matrix elements in Eq. (\ref{cf}). 
We now assume that the large-distance asymptotics of $C(l)$ is
determined by the lowest excited state in the $S^z=-1$ sector. The
difference in wave number between this state and the ground state then
gives the incommensurability $q_{\min }$.

For a finite system, in order to represent a continuum of wave vectors, we
minimize the energy in the $S^z=-1$ sector $E_{1}(K_{1},\Phi )$, as a
function of flux and allowed discrete wave vectors for this flux (see Eq. (%
\ref{ks})). The wave vector of the excitation is taken as:

\begin{equation}
q_{\min }=\theta =K_{1}(\Phi _{\min })-K_{0}(\Phi _{\min }),  \label{qmin}
\end{equation}
where $\Phi _{\min }$ is the flux which minimizes $E_{1}(K_{1},\Phi )$, and $%
K_{i}(\Phi )$ is the wave vector of the state of lowest energy in the $%
S^z=-1$ sector at flux $\Phi $. The wave vector $q_{\min }$ gives the
period of the oscillations of the spin correlations and corresponds to
the pitch angle $\theta $ of the classical spiral density wave \cite{bur}.

We have also investigated the energy gap of the spectrum. For a finite
system we evaluate it as:

\begin{equation}
\Delta _{g}=E_{1}(K_{1},\Phi _{\min })-E_{0}(K_{0},\Phi _{\min }^{\prime }).
\label{gap}
\end{equation}
Here $\Phi _{\min }^{\prime }$ is the flux which minimizes the ground-state
energy $E_{0}$. There are alternative expressions to $q_{\min }$ and $\Delta
_{g}$ which converge to the same value in the thermodynamic limit. Our
experience suggests that the ones we chose have the fastest convergence.

To test the method, we have studied a dimerized half filled system of non
interacting spinless fermions. This model is the fermionic version of an $XY$
model plus additional interactions:

\bea
H_{t}&=&-\sum_{l=1}^{4}t_{l}\sum_{i}(c_{i+l}^{\dagger }c_{i}+{\rm H.c.})\nn
&&+\frac{V}{2}\sum_{i}(-1)^{i}(c_{i+1}^{\dagger }c_{i}+{\rm H.c.}).  \label{h2}
\eea
We have taken the parameters $t_{1}=1$, $t_{2}=0.4$, $t_{3}=0$, $t_{4}=-0.2$%
, $V=0.3$, in such a way that the upper (empty) band has minima at
incommensurate wave vectors $q_{\min }=\pm 0.2356$ $\pi $, and the lower
(full) band has its maximum at $q=0$. There is an indirect gap $\Delta
_{g}=0.12477$.

\begin{figure}
\narrowtext
\epsfxsize=3.0truein
\vbox{\hskip 0.05truein \epsffile{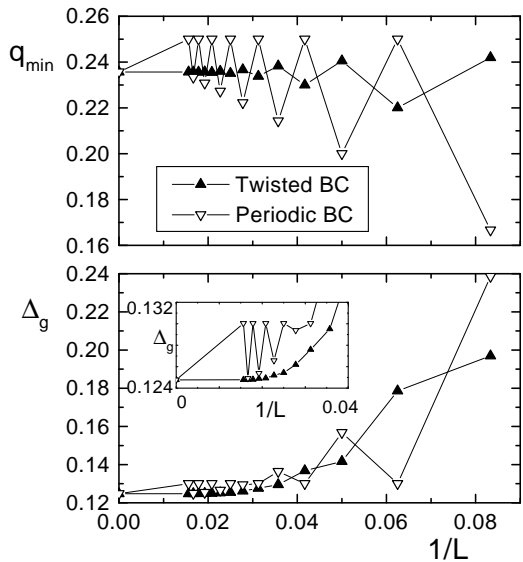}}
\medskip
\caption{Incommensurate wave vector (top) and gap (bottom) as a function
of system size, obtained using periodic and twisted boundary conditions, for
the toy model Eq. (8).}
\end{figure}

In Fig. 2, we compare the results for $q_{\min }$ and
$\Delta _{g}$ in 
finite systems of $L$ sites, with $L/4$ integer, and $12\leqslant L\leqslant
64$, between PBC and TBC. The gap is calculated as $\Delta _{g}=E_{1}(\Phi
_{1})+E_{-1}(\Phi _{-1})-2E_{0}(\Phi _{0})$, where $E_{i}(\Phi _{i})$ is the
ground-state energy for $i$ added particles. For TBC the fluxes $\Phi _{i}$
are those which minimize $E_{i}(\Phi _{i})$, while for PBC $\Phi _{i}=0$.
Since $q_{\min }$ is near $\pi /4$, and the latter is one of the allowed
wave vectors of PBC for $L$ multiple of 8, small periodic systems with $L/8$
integer have the minimum energy for one added particle, when this particle
has wave vector  $\pm \pi /4$. As long as $2\pi /L$ is larger than $\pi
/4-|q_{\min }|$ (small systems), the results for PBC are better if $L$ is
multiple of 8. However, the oscillations with increasing $L$ for PBC make a
finite-size scaling difficult. Although $q_{\min }$ also oscillates with $L$
for TBC, the oscillations are smaller and the convergence to the
thermodynamic limit is much faster. Using TBC, for $L\eqslantgtr 52$, the
error in $q_{\min }$ and $\Delta _{g}$ are below 0.01\%. For the maximum
size of the system used in our Lanczos diagonalization of Eq. (\ref{h}) ($%
L=24$), the error in $q_{\min }$ is $\sim 2\%$ and that of $\Delta _{g}$ is
of the order of 10\%.

\section{Results}

\subsubsection{a) Isotropic case}

The incommensurate spin correlations and spin gap for $\Delta =1$ and $%
J_{2}/J_{1}<3$ have been studied previously by DMRG \cite{bur,whi}. In
this section, we compare our results with these ones, and extend the
study to $3\leq J_{2}/J_{1}\leq 30,$ a region which is very difficult
to reach with other methods. 

We have used a linear extrapolation in $1/L$ of the data for the angle $%
\theta =q_{\min }$ for $L=12,16,20,$ and $24$. We have chosen $L/4$ to
be an integer in order to avoid frustration of antiferromagnetic
interactions for large $J_{2}$. A quadratic fit gives smaller values
of $\theta $, which underestimate the DMRG\ results. The comparison of
available DMRG results and those obtained using TBC as described in
the previous section is included in Fig. 3. For $J_{2}/J_{1}<0.7$ we
do not obtain any incommensurability. This is probably a
finite size effect, since for $J_{2}/J_{1}=0.7$, we obtain $\theta =\pi$
for $L=12$, but incommensurate values of $\theta $ for $L>12$. We have
disregarded the value for $L=12$ in the extrapolation when
$J_{2}/J_{1}=0.7$. 

\begin{figure}
\narrowtext
\epsfxsize=3.0truein
\vbox{\hskip 0.05truein \epsffile{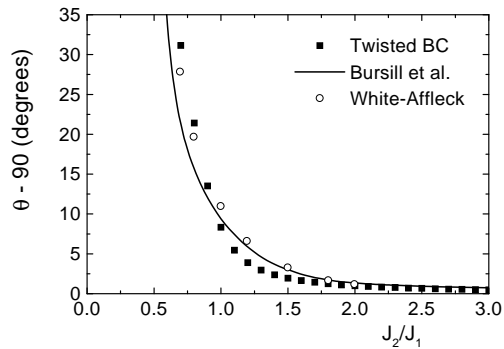}}
\medskip
\caption{Incommensurate angle as a function of $J_{2}/J_{1}$ for $
\Delta =1$, obtained using twisted boundary conditions. The DMRG results of
Refs. 7,8 are also shown.}
\end{figure}

For $J_{2}/J_{1}\leq 1$ and $J_{2}/J_{1}\geq 1.8$ our results are in better
agreement with those of White and Affleck \cite{whi} than with those of
Bursill {\it et al }\cite{bur}. In the remaining region both DMRG\ results
are very similar. In general, the difference between the nearest of the
results of the three calculations for $\theta -90^{\circ }$ is of the order
of $20\%$. For example for $J_{2}/J_{1}=0.7$ our results and those of Ref.%
\cite{whi,bur} are respectively 31$^{\circ }$, 28$^{\circ }$ and 21$^{\circ }
$. For $J_{2}/J_{1}=2$ the corresponding values are 1.00$^{\circ }$, 1.19$%
^{\circ }$, and 1.36$^{\circ }$ . In general, comparison with DMRG results
and the difference between different extrapolation methods suggest that the
error in  $\theta -90^{\circ }$ using TBC is roughly of the order of 20\%.
In view of the simplicity of our method compared to DMRG\ calculations, we
believe that our results are satisfactory. Note that to detect an
incommensurability of 1$^{\circ }$ without using TBC, the size of the system
should be of the order of $L\sim 360$ !

\begin{figure}
\narrowtext
\epsfxsize=3.0truein
\vbox{\hskip 0.05truein \epsffile{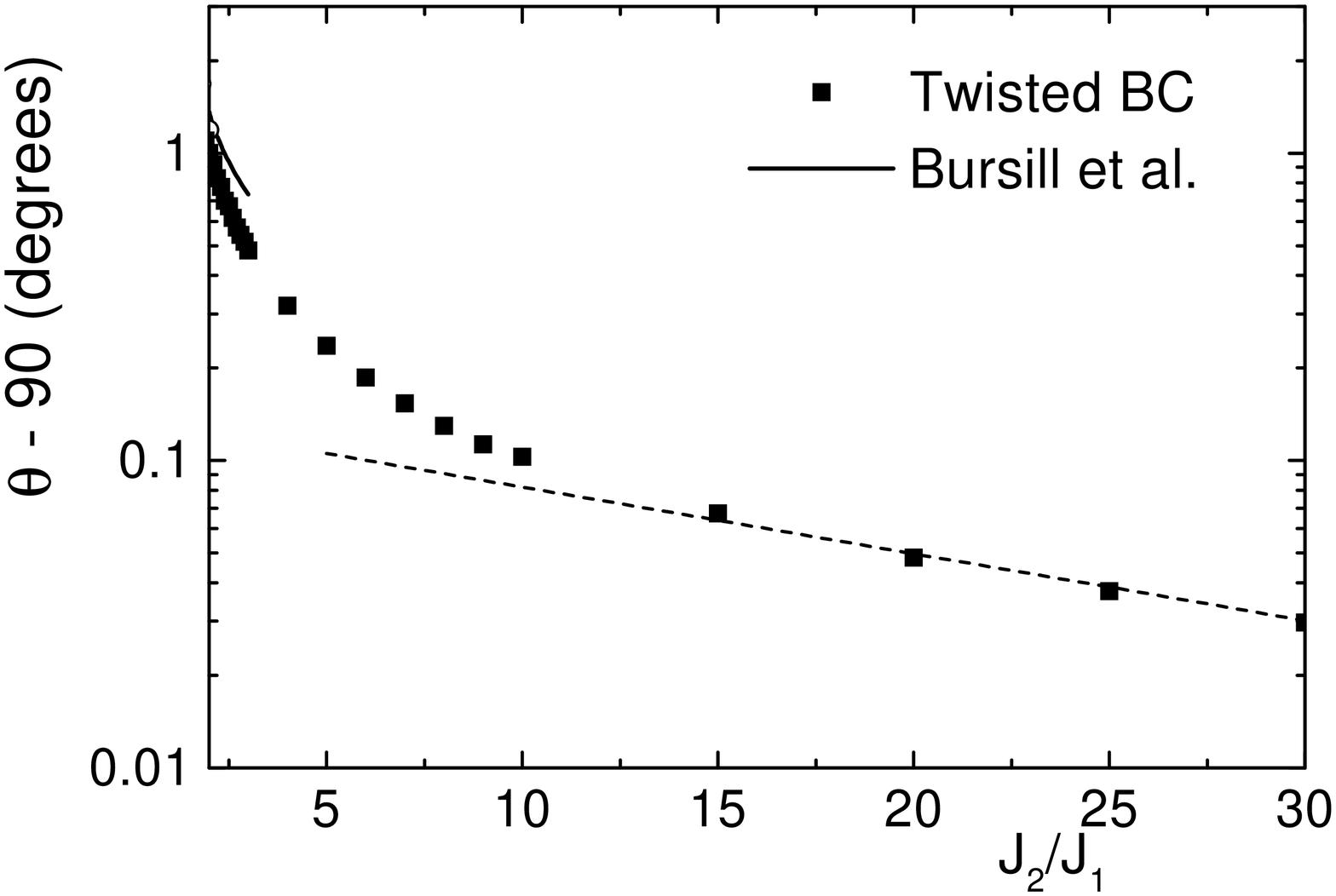}}
\medskip
\caption{Same as Fig. 2 for $J_{2}/J_{1}>2$ in logarithmic scale.
Dashed line is the function $e^{-2-0.05J_{2}/J_{1}}$.}
\end{figure}

In Fig. 4, we show the angle as a function of $J_{2}/J_{1},$ for $2\leq
J_{2}/J_{1}\leq 30$ on a logarithmic scale. In this region the
incommensurability is very small, and therefore, as explained above, very
hard to obtain by alternative numerical methods. For large $J_{2}/J_{1}$,
the deviation of the angle from $\pi /2$ is expected to be of the form
\cite{whi,NGE} 

\begin{equation}
\theta -90^{\circ }=\exp (-a-bJ_{2}/J_{1})
\end{equation}
A linear fit of $\ln (\theta -90^{\circ })$ as a function of
$J_{2}/J_{1}$ in the interval [15,30] gives within a few percent 
$a=2 $, and $b=1/20.$

\begin{figure}
\narrowtext
\epsfxsize=3.0truein
\vbox{\hskip 0.05truein \epsffile{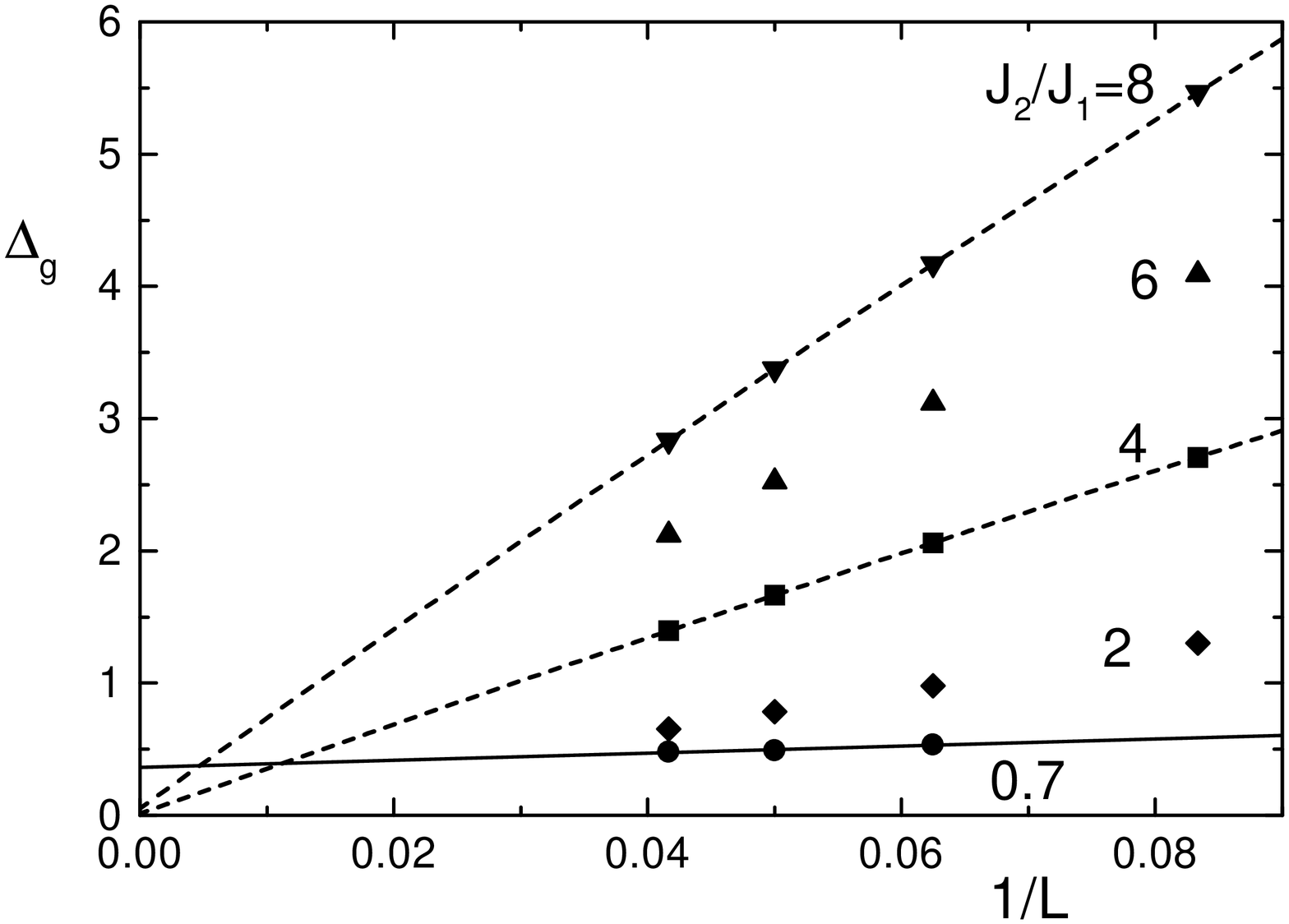}}
\medskip
\caption{Size dependence of the gap for $\Delta =1$ and several values
of $J_{2}/J_{1}$. Dashed (full) lines are quadratic (linear) fits of the
data.}
\end{figure}

We have also used \r{gap} to study the gap $\Delta _{g}$. Due to the
smallness of the gap, the finite-size effects for the
relatively small system sizes we consider are too large to allow us to obtain 
reliable values for $\Delta _{g}$. Fig. 5 shows the size dependence of 
$\Delta _{g}$. From the difference between linear and quadratic
extrapolation, we estimate the error in the extrapolated gap to be of
the order of $0.1J_{1}$, while for any value of $J_{2}/J_{1}$, $\Delta
_{g}<0.5 J_1$. Within this error, our results agree with those
reported by White and Affleck \cite {whi}.

\subsubsection{b) Anisotropic case}

In \cite{NGE} the zigzag ladder was studied by means of a field theory
approach in the regime $J_2\gg J_1$. A mechanism for generating
incommensurabilities was identified and analysed quantitatively for
the case of two coupled XX chains ($\Delta=0$). It was found that spin
correlations exhibit a very slow power-law decay and are incommensurate
\begin{equation}
\langle S_1^+(x) S_j^-(0)\rangle\sim \frac{(-1)^{x/a_0}}{|x|^{1/4}}\
\exp\left[-i\kappa x/a_0\right]
\label{asymp}
\end{equation}
where $j=1,2$ and the deviation of the pitch angle from $\pi$ is
$\kappa\propto (J_1/J_2)^2$. The analysis of \cite{NGE} also implies
the existence of local magnetisation currents around the elementary
triangular plaquettes of the ladder.
The findings of \cite{NGE} were questioned in \cite{kab}, where the
squares of the local magnetisation currents were computed numerically
and found to decrease with system size for open chains of up to 16
sites. 

\begin{figure}
\narrowtext
\epsfxsize=3.0truein
\vbox{\hskip 0.05truein \epsffile{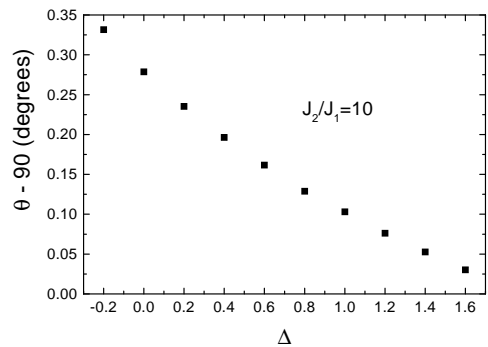}}
\medskip
\caption{Incommensurate angle as a function of $\Delta $ for $%
J_{2}/J_{1}=10.$}
\end{figure}

In an attempt to resolve this controversy we have set $J_{2}/J_{1}=10$
and studied the variation of the incommensurate angle with the
anisotropy parameter $\Delta $. As shown in Fig. 6, we find that
$\theta -90^\circ$ increases considerably as $ \Delta $ is decreased
from the isotropic case $\Delta =1$. This is in agreement with the
field-theory prediction of \cite{NGE}.

We futhermore have determined the dependence of the incommensurability
on $J_2/J_1$. The results are shown in Fig. 7.
For large values of $J_2/J_1$ our numerical results are well fitted by
\be
\theta-90^\circ =2.785^\circ \frac{J_1}{J_2}\ .
\label{linear}
\ee
This is in disagreement with the prediction of \cite{NGE}.

The disagreement between the predictions of \cite{NGE} 
and the numerical results \r{linear} and \cite{kab} could
either be due to a defect in the mean-field solution of \cite{NGE} or
be an artifact of the limited system sizes used in the numerical
computations. In fact, the analysis of \cite{NGE} predicts a gapless
phase at $\Delta=0$ so that it is conceivable that numerical results
for small clusters are plagued by finite-size effects.

\begin{figure}
\narrowtext
\epsfxsize=3.0truein
\vbox{\hskip 0.05truein \epsffile{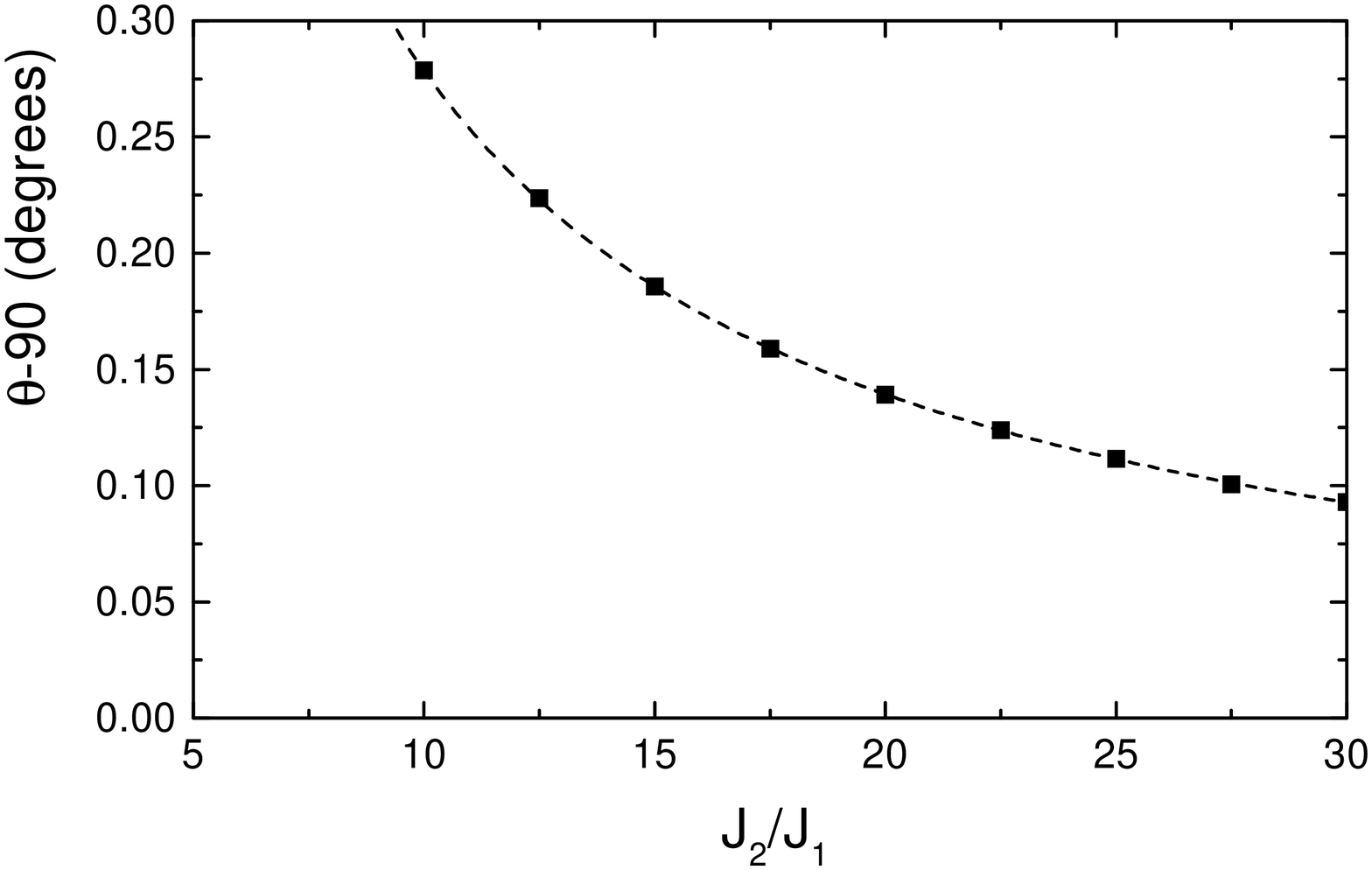}}
\medskip
\caption{Incommensurate angle as a function of $J_2/J_1$ for $\Delta=0$.
Dashed line is the function $2.785^\circ J_1/J_2$.}
\end{figure}

There is indeed evidence suggesting that finite-size effects are
still significant for $L=24$. We find that the ground state in the
$S^z=1$ sector has a lower energy than the first excited state in the
$S^z=0$ sector for $L=24$ and ${J_2}{\stackrel{\textstyle >}{\sim}}2J_1$. 
On the other hand, the DMRG studies of \cite{whi} show that in the
isotropic case and for long lattices the two lowest levels are both in
the $S^z=0$ sector (degenerate ground states corresponding to
different signs of the dimerisation). 

We note that the presence of such finite-size effects does not
necessary imply that the extrapolated results for the
incommensurability are incorrect.
In order to resolve this issue it is necessary to study significantly
longer lattices.

\section{Summary and discussion}

We have calculated the incommensurate wave number in the
next-nearest-neighbor Heisenberg model with anisotropy $\Delta $, by
exact diagonalization of rings of up to 24 sites. We have used twisted
boundary conditions and assumed that the incommensurate spin
fluctuations for $|\Delta |\eqslantless 1$ are determined by the
lowest excited state for total spin projection $S^z=\pm 1 $. 

The method is able to detect incommensurate angles $\theta \sim
0.03^{\circ }$. This corresponds to a wave length of the order of 10 000
sites and is impossible to detect by using alternative numerical methods.
However, for certain parameters ($J_{2}/J_{1}<0.7$ in the model), our
method is unable to detect the incommensurability although it is
rather large. 
On the other hand, the method does not predict incommensurabilities in
cases where it is known that none exist.
Also, in general, the extrapolated vales of $\theta $ seem to be
underestimations, as compared with known DMRG results. This is also
the case for the toy model (Eq. (\ref{h2})), represented in Fig. 2,
where a linear extrapolation gives an underestimation of $q_{\min }$
by $\sim 30\%$ if the results are limited to 24 sites.

The advantage of using TBC for facilitating a finite-size scaling
analysis has been noted previously in e.g. \cite{fath}, but their use
for detecting incommensurabilities is to the best of our knowledge novel.

In spite of the limitations of the size of the cluster, the values of $%
\theta $ obtained with our method are in reasonable agreement with known
DMRG results in the isotropic case. For this case, we have also studied the
region $J_{2}/J_{1}>3$, which is very difficult to reach by alternative
methods. For sufficiently large $J_{2}/J_{1}$, $\theta -\pi /2$ decays as $%
\exp (-b$ $J_{2}/J_{1})$ as predicted by field theory \cite{whi}. We obtain
that the constant $b\sim 1/20$.

In the anisotropic case, we obtain that $\theta $ increases with
decreasing $\Delta $, in agreement with Ref. \cite{NGE}. However, we
find a {\sl linear} dependence of $\theta-\pi/2$ on $J_1/J_2$ for
$J_2\gg J_1$, in contrast to the quadratic behaviour predicted in
\cite{NGE}. This discrepancy may be due to finite-size effects.

In summary, we have shown that incommensurabilities can be detected by
diagonalizing finite-size clusters with TBC. The main advantage of
the method is that the dependence of the incommensurability on system
size is very smooth and allows extrapolation from results for
relatively short chains.

It would be very interesting to implement our TBC method in a DMRG
algorithm and study the anisotropic zigzag chain for much larger sizes.

\section*{Acknowledgements}

We are grateful to A. Lopez for important discussions and to
R. Bursill for providing us with the numerical data of
Ref. \cite{bur}. C. D. B. is supported by CONICET,
Argentina. A. A. A. is partially supported by CONICET. F.H.L.E. is
supported by the EPSRC under grant AF/98/1081. This work was sponsored
by the cooperation agreement between British Council and Fundaci\'{o}n
Antorchas, PICT 03-00121-02153 of ANPCyT and PIP 4952/96 of CONICET.

\end{document}